
\magnification \magstep1
\raggedbottom
\openup 4\jot
\voffset6truemm
\headline={\ifnum\pageno=1\hfill\else
\hfill {\it Quantization and Regularization in Perturbative
Quantum Cosmology} \hfill \fi}
\centerline {\bf QUANTIZATION AND REGULARIZATION IN}
\centerline {\bf PERTURBATIVE QUANTUM COSMOLOGY}
\vskip 1cm
\centerline {\bf Giampiero Esposito}
\vskip 1cm
\centerline {\it International Centre for Theoretical Physics,}
\centerline {\it Strada Costiera 11, 34014 Trieste, Italy}
\centerline {\it Scuola Internazionale Superiore di Studi Avanzati,}
\centerline {\it Via Beirut 2-4, 34013 Trieste, Italy}
\vskip 1cm
\noindent
{\bf Abstract.} Reduction to physical degrees of freedom before
quantization leads to predictions for one-loop amplitudes in quantum
cosmology in the presence of boundaries which disagree with the
results obtained from Faddeev-Popov theory and boundary-counterterms
technique. However, the mode-by-mode analysis of eigenvalue
equations for gauge modes and ghost fields remains a very
difficult problem. Hence the equivalence or inequivalence of various
quantization and regularization techniques cannot be easily proved.
\vskip 1cm
\leftline {PACS 03.70.+k - Theory of quantized fields.}
\leftline {PACS 04.60.+n - Quantum theory of gravitation.}
\vskip 1cm
\centerline {SISSA Ref. 4/93/A (January 1993)}
\vskip 10cm
In the last few years, much work has been done on perturbative
properties of physical theories in the presence of boundaries
within the framework of one-loop quantum cosmology [1]. The
corresponding problems are as follows.

(i) {\it Choice of locally supersymmetric boundary conditions} [1]:
they involve the normal to the boundary and the field for spin
${1\over 2}$, the normal to the boundary and the
spin-${3\over 2}$ potential for gravitinos, Dirichlet conditions
for real scalar fields, magnetic or electric field for electromagnetism,
mixed boundary conditions for the four-metric of the gravitational
field (and in particular Dirichlet conditions on the perturbed
three-metric).

(ii) {\it Quantization techniques}: one-loop amplitudes can be
evaluated by first reducing the classical theory to the physical
degrees of freedom (hereafter referred to as PDF) by choice of
gauge and then quantizing, or by using the gauge-averaging method
of Faddeev and Popov, or by applying the extended-phase-space
Hamiltonian path integral of Batalin, Fradkin and Vilkovisky [1].

(iii) {\it Regularization techniques}: the generalized Riemann
zeta-function and its regularized $\zeta(0)$ value
(which yields both the scaling of the one-loop prefactor and the
one-loop divergences of physical theories) can be
obtained by studying the eigenvalue equations obeyed by perturbative
modes, once the corresponding degeneracies and boundary conditions
are known, or by using geometrical formulae for one-loop
counterterms which generalize well-known results for scalar fields,
but make no use of mode-by-mode eigenvalue conditions and degeneracies.

It turns out that one-loop quantum cosmology may add further evidence
in favour of different approaches to quantizing gauge theories being
inequivalent. Studying flat Euclidean backgrounds bounded by a
three-sphere, for electromagnetism the PDF method yields
$\zeta(0)=-{77\over 180}$ and $\zeta(0)={13\over 180}$ in the magnetic
and electric cases respectively [1], whereas the {\it indirect}
Faddeev-Popov method (i.e. when one-loop amplitudes are expressed
using the boundary-counterterms technique and evaluating the various
coefficients as in ref.[2]) is found to yield
$\zeta(0)=-{38\over 45}$ in both cases [1,3]. For $N=1$ supergravity,
the PDF method yields {\it partial} cancellations between spin $2$
and spin ${3\over 2}$ [1], whereas the {\it indirect} Faddeev-Popov
method yields a one-loop amplitude which is even more divergent than
in the pure-gravity case [2]. Finally, for pure gravity, the PDF
method yields $\zeta(0)=-{278\over 45}$ in the Dirichlet case,
whereas the {\it indirect} Faddeev-Popov method yields
$\zeta(0)=-{803\over 45}$ [1,2]. Moreover, within the PDF approach, it
is possible to set to zero on $S^{3}$ the linearized magnetic
curvature. This yields a well-defined one-loop calculation, and the
corresponding $\zeta(0)$ value is ${112\over 45}$ [1]. By
contrast, using the Faddeev-Popov formula, magnetic boundary conditions
for pure gravity are ruled out [2].

It is therefore necessary to get a better understanding of the
manifestly gauge-invariant formulae for one-loop amplitudes
used so far in the literature, by performing a mode-by-mode
analysis of the eigenvalue equations, rather than relying on
general formulae which contain no explicit information about
degeneracies and eigenvalue conditions. As shown in ref.[1], this
detailed analysis can be attempted for vacuum Maxwell theory studied at
one-loop about flat Euclidean backgrounds bounded by a three-sphere,
recently considered in perturbative quantum cosmology.
Working within the Faddeev-Popov formalism and making a $3+1$
split of the vector potential, the full $\zeta(0)$ value takes
into account the contribution of the physical degrees of freedom,
i.e. the transverse part $A^{(T)}$ of the vector potential, the
gauge modes, i.e. the $A_{0}$ component and the longitudinal part
$A^{(L)}$ of the vector potential, and the ghost action.
Interestingly, a gauge-averaging term can be found such that the
contributions to $\zeta(0)$
of physical degrees of freedom and of decoupled mode
for $A_{0}$ add up to $-{61\over 90}$ both in the electric and in
the magnetic case. However, remaining modes for $A_{0}$ and
$A^{(L)}$ are always found to obey a coupled system of second-order
ordinary differential equations. This system has been solved exactly
[1], but unfortunately the power series appearing in its solution
are not (obviously) related to well-known special functions. The
corresponding asymptotic analysis (i.e. at large values of the
eigenvalues) is therefore much harder, and remains a stimulating
challenge for applied mathematicians and theoretical physicists.
It also turns out that, in the presence of boundaries, gauge modes
should remain coupled, for us to be able to find a linear,
elliptic second-order operator corresponding to the ghost action.
Since the difficulties concerning gauge modes and ghost fields are
technical in nature and not completely unfamiliar
(i.e. systems of ordinary differential equations, fourth-order
algebraic equations, finite parts of diverging series), there is
hope that the research initiated in ref.[1]
(see corrections in the Appendix) will shed new light on
one-loop properties of physical theories in the presence of
boundaries.
\vskip 1cm
\leftline {APPENDIX}
\vskip 1cm
Since we rely on ref.[1], we should correct the mistakes appearing
therein. They are as follows. The comments (1), (2) and (3) following
eq. (2.3.22) should read: There are secondary constraints, i.e.
$q_{2} \approx 0$, and all constraints are second-class. A similar
correction should be made on page 37 (see ref.[4]). On page 64, the
exponential $e^{r\over 2M}$ on the right-hand side of eqs.
(3.3.14)-(3.3.15) should read $e^{r\over 4M}$. On page 170, the
second line following eq. (6.5.43) should read: ..., {\it the partial
effect} of the ghost field, and {\it the effect} of the $R_{1}$-mode.
Indeed, there are additional contributions since the gauge-averaging
functional of eq. (6.5.9) is not the Lorentz functional
${ }^{(4)}\nabla^{\mu}A_{\mu}$.
\vskip 1cm
\leftline {REFERENCES}
\vskip 1cm
\item {[1]}
G. ESPOSITO: {\it Quantum Gravity, Quantum Cosmology and Lorentzian
Geometries}, Lecture Notes in Physics, New Series m: Monographs,
Volume m12 (Springer-Verlag, Berlin, 1992).
\item {[2]}
S. POLETTI: {\it Phys. Lett.}, B {\bf 249}, 249 (1990).
\item {[3]}
I. G. MOSS and S. POLETTI: {\it Phys. Lett.}, B {\bf 245}, 355 (1990).
\item {[4]}
A. CORICHI: {\it J. Math. Phys.}, {\bf 33}, 4066 (1992).
\bye